\documentclass[conference]{IEEEtran}
\IEEEoverridecommandlockouts
\usepackage{cite}
\usepackage{amsmath,amssymb,amsfonts}
\usepackage{graphicx}
\usepackage{textcomp}
\usepackage{xcolor}
\def\BibTeX{{\rm B\kern-.05em{\sc i\kern-.025em b}\kern-.08em
    T\kern-.1667em\lower.7ex\hbox{E}\kern-.125emX}}
\usepackage[utf8]{inputenc}
\usepackage{csquotes}
\usepackage{times}
\usepackage{url}
\usepackage{array}
\usepackage{multirow}
\usepackage{algorithm,algcompatible}
\usepackage{caption}
\usepackage{subfigure}
\usepackage{scalerel}

\usepackage{tikz}  
\algnewcommand\INPUT{\item[\textbf{Input:}]}%
\algnewcommand\OUTPUT{\item[\textbf{Output:}]}%

\usetikzlibrary{svg.path}
\definecolor{orcidlogocol}{HTML}{A6CE39}
\tikzset{
  orcidlogo/.pic={
    \fill[orcidlogocol] svg{M256,128c0,70.7-57.3,128-128,128C57.3,256,0,198.7,0,128C0,57.3,57.3,0,128,0C198.7,0,256,57.3,256,128z};
    \fill[white] svg{M86.3,186.2H70.9V79.1h15.4v48.4V186.2z}
                 svg{M108.9,79.1h41.6c39.6,0,57,28.3,57,53.6c0,27.5-21.5,53.6-56.8,53.6h-41.8V79.1z M124.3,172.4h24.5c34.9,0,42.9-26.5,42.9-39.7c0-21.5-13.7-39.7-43.7-39.7h-23.7V172.4z}
                 svg{M88.7,56.8c0,5.5-4.5,10.1-10.1,10.1c-5.6,0-10.1-4.6-10.1-10.1c0-5.6,4.5-10.1,10.1-10.1C84.2,46.7,88.7,51.3,88.7,56.8z};
  }
}
\newcommand\orcidicon[1]{\href{https://orcid.org/#1}{\mbox{\scalerel*{
\begin{tikzpicture}[xscale=0.05,yscale=-0.05,transform shape]
\pic{orcidlogo};
\end{tikzpicture}
}{|}}}}

\usepackage{hyperref} 

\begin{document}

\title{Disturbances in Influence of a Shepherding Agent is More Impactful than Sensorial Noise During Swarm Guidance\\
}
\author{\IEEEauthorblockN{Hung The Nguyen~$^{\orcidicon{0000-0002-0938-0930}}$, Matthew Garratt~$^{\orcidicon{0000-0003-0222-430X}}$, Lam Thu Bui~$^{\orcidicon{0000-0002-9812-4970}}$, Hussein Abbass~$^{\orcidicon{0000-0002-8837-0748}}$}\\
\IEEEauthorblockA{\IEEEauthorrefmark{1}\textit{School of Engineering and Information Technology, UNSW-Canberra,  University of New South Wales}\\
Canberra, Australia.}
\IEEEauthorblockA{\IEEEauthorrefmark{2}\textit{Department of Information Technology, Le Quy Don Technical University
Hanoi, Vietnam}}
}

\maketitle
\IEEEoverridecommandlockouts
\IEEEpubid{\makebox[\columnwidth]
{978-1-7281-2547-3/20/\$31.00~\copyright2020 IEEE \hfill} 
\hspace{\columnsep}\makebox[\columnwidth]{ }}
\IEEEpubidadjcol
\begin{abstract}  
The guidance of a large swarm is a challenging control problem. Shepherding offers one approach to guide a large swarm using a few shepherding agents (sheepdogs). While noise is an inherent characteristic in many real-world problems, the impact of noise on shepherding is not a well-studied problem. We study two forms of noise. First, we evaluate noise in the sensorial information received by the shepherd about the location of sheep. Second, we evaluate noise in the ability of the sheepdog to influence sheep due to disturbance forces occurring during actuation. We study both types of noise in this paper, and investigate the performance of Str\"{o}mbom\textquoteright s approach under these actuation and perception noises. To ensure that the parameterisation of the algorithm creates a stable performance, we need to run a large number of simulations, while increasing the number of random episodes until stability is achieved. We then systematically study the impact of sensorial and actuation noise on performance. Str\"{o}mbom\textquoteright s approach is found to be more sensitive to actuation noise than perception noise. This implies that it is more important for the shepherding agent to influence the sheep more accurately by reducing actuation noise than attempting to reduce noise in its sensors. Moreover, different levels of noise required different parameterisation for the shepherding agent, where the threshold needed by an agent to decide whether or not to collect astray sheep is different for different noise levels.
\end{abstract}



\section{Introduction}

The research area of bio-inspired swarm control has attracted diverse ideas and perspectives from a variety of research areas, including control theory, biology, and artificial intelligence~\cite{martinez2007motion}. These bio-inspired approaches provide valuable insights into designing multi-agent systems and/or robotic swarms. In these systems, studies aim to address the question of how to control a swarm of individual agents based on natural interactions occurring among the agents and between the agents and their operating environment~\cite{carelli2006centralized,oh2017bio}. 

The shepherding problem is inspired by sheep-herding in agriculture wherein a single or multiple shepherds or sheepdogs are used to guide a large group of sheep. Shepherding is considered as a \emph{guided-flocking behaviour} when one or multiple external agents acting as shepherds, drive a swarm of individual agents, called \emph{flocking} or \emph{sheep agents} towards a given target. The herding idea has been applied to the field of multi-agent systems and swarm robotics~\cite{strombom2018robot}. There are many applications~\cite{strombom2018robot,nalepka2019practical} of the shepherding problem within these fields such as herding living animals such as driving a large group of bird or sheep in a field area, assisting in controlling human crowd activities, cleaning environmental hazards such as oil-spills, or guiding cells to fix tissues in internal medicine~\cite{cohen2014galvanotactic}.

In recent years, a heuristic approach, which is proposed by Str\"{o}mbom et al.~\cite{strombom2014solving, strombom2018robot}, has addressed the shepherding problem effectively. In this approach, two main behaviours are used: \emph{collecting} and \emph{driving}, which are used to explain the interaction between a shepherd considered as one intelligent agent and a swarm of sheep being treated as autonomous agents. The collecting behaviour aims to maintain the entire sheep grouped within a connected network, while the driving behaviour enables guidance of the group/swarm of sheep towards a goal. The Str\"{o}mbom approach shows good shepherding performance in successfully collecting and driving a large number of sheep towards the given target. However, Str\"{o}mbom et al. evaluate their approach in an ideal environment possessing an unrealistically low noise level for both the shepherd and the sheep. In practice, the shepherd might face various sources of significant noise associated with the response of the sheep to the influence of the sheepdog, called the \emph{actuation noise}, and noise coming from the sensing ability of the shepherd, called the \emph{perception noise}. Under extreme weather conditions, or obstacles, the sheep or autonomous agents might move very imprecisely in response to shepherding commands, and the shepherd can inadequately sense the position of the sheep. These errors may lead to poor performance of the shepherd. In the literature, there are insufficient attempts to understand the impact of these errors on the overall performance of the shepherding in the successful completion of the task.

In this paper, we first investigate the level of evaluations sufficient to achieve stability in performance in the Str\"{o}mbom approach~\cite{strombom2014solving}. After that, we study the performance of the model under increasing actuation and perception noises. Furthermore, we identify appropriate thresholds of switching between the collecting and driving behaviours, called the collecting frequency, in order to improve the performance of the shepherd. To identify these thresholds, we systematically decrease and increase the threshold value used in the Str\"{o}mbom method. Our experiments are conducted in the same simulation environment as that introduced in the Str\"{o}mbom approach. The results from the experiments show that the performance of the shepherd is more sensitive to perception noise than actuation noise. Moreover, a guiding set of appropriate thresholds are identified which should help improve the shepherding efficiency by adapting the switch between collecting and driving behaviours to suit the amount of perception and actuation noises present.

The remainder of the paper is organized as follows. In Section~\ref{sect:relatedwork}, we provide a brief review of the research focused on the shepherding problem in order to identify a gap in the evaluation of shepherding performance under noise conditions. Following this section, we formally define shepherding using an appropriate notional system and a corresponding mathematical objective in Section~\ref{sect:shepherding}. The proposed evaluation framework is introduced in Section~\ref{sect:framework}. The framework is conducted in a simulated shepherding task in Section~\ref{sect:experiment}. Section~\ref{sect:results} presents the results of the framework. Conclusions are drawn in Section~\ref{sect:conclusions}, followed by a discussion on future work.

\section{Related work}\label{sect:relatedwork}

In nature, the behaviours of flocking of birds, herding  of  land  animals,  or schooling of fish can be widely seen~\cite{reynolds1987flocks}. In these behaviours, a large number of individual agents will be influenced by one controlling agent in order to successfully achieve different goals such as finding food or foraging. Studying the various kinds of swarm behaviours in nature can greatly assist in the design of distributed and coordinated control methods for robotic swarms or multi-agent systems~\cite{kennedy1995particle,eberhart1995new}.

Early research on the shepherding problem was carried out by Schultz et al. ~\cite{schultz1996roboshepherd}. In this research, the authors use genetic algorithms to learn rules for a shepherd or sheep-dog agent so that it might drive a swarm of sheep towards a desired target. Lien et al.~\cite{lien2004shepherding} conducted experiments in order to simulate four main behaviours: herding, covering, patrolling, and collecting. The combination of these four behaviours for the shepherd shows effective shepherding strategies. However, both pieces of research are more suitable for driving a small number of sheep (less than 40)~\cite{bennett2012comparative}.

Towards guiding a larger number of sheep agents (more than 40), Str\"{o}mbom et al.~\cite{strombom2014solving,strombom2018robot} used a heuristic approach to choose an appropriate threshold to switch between collecting and driving behaviours in order to guide the entire sheep. The approach is promising, enabling the shepherd to guide up to 300 sheep effectively. Adopting Str\"{o}mbom approach, some other research has attempted to use artificial intelligence methods, such as reinforcement learning~\cite{nguyen2019deep}, apprenticeship learning~\cite{nguyen2019apprenticeship}, and machine education~\cite{gee2019transparent,clayton2019machine}. However, in both the Str\"{o}mbom method and the research adopting the approach of Str\"{o}mbom, the shepherd works in an ideal environment with just a small amount of noise added to the shepherd and the sheep to avoid deadlocks. In practice, the operating environment of these agents might include various significant noise sources impacting on the performance of the shepherd. One source of noise comes from responses and behaviours of the sheep that deviate from the intent of the sheepdog due to differences between intent and actuation, or actuation and response, we group them under actuation noise. Another source of noise is the sensing ability of the shepherd, called the perception noise. To date, there has not been any published work on contrasting these noises on the performance of a shepherd for swarm guidance.

\section{Shepherding Problem}\label{sect:shepherding}

In the shepherding problem, Str\"{o}mbom et al.~\cite{strombom2014solving} introduce a heuristic approach in which the movement of sheep is computed, and from there an effective control strategy is created for the shepherd. In this paper, the Str\"{o}mbom et al.~\cite{strombom2014solving} approach is described by providing the notations as well as the mechanism that we will use later in the experimental design.

The operating environment of the shepherding problem is a 2-D square paddock having length of $L$. In this environment, two kinds of agents, which are a set of sheep (called influenced agents) $\Pi = \{ \pi_1, \dots, \pi_i, \dots, \pi_N \}$, and a set of shepherds (called influencing agents) $B = \{ \beta_1, \dots, \beta_j, \dots, \beta_M \}$, are initialized. There are three main behaviours for each shepherd, and four basic behaviours for each sheep at a time step $t$. These behaviours are shown as below.

\begin{enumerate}
\item {For shepherd $\beta_j$:
\begin{itemize}
\item \emph{Driving behaviour} $\sigma_1$: When all sheep are collected in a cluster, i.e. all the distances from the observed sheep to the center of sheep\textquoteright s mass are lower than a threshold $f(N)$ calculated in Equation~\ref{eq:ShepherdBehaviorSelectionThresholdEquation}, a normalized force vector, $F^t_{\beta_jcd}$, is applied for the shepherd as a velocity vector in order to reach a driving point. This point is located behind the sheep\textquoteright s mass on the line drawn from the center of the sheep\textquoteright s mass and the target position. The distance from the center of the mass to the driving point is $R_{\pi\pi}\times \sqrt{N}$.

\begin{equation}\label{eq:ShepherdBehaviorSelectionThresholdEquation}
f(N) = R_{\pi\pi} N^\frac{2}{3}
\end{equation}

where $R_{\pi\pi}$ is the sensing range among sheep.

\item \emph{Collecting behaviour} $\sigma_2$: When a sheep is deemed to have gone astray from the others i.e. the distance from the sheep to the center of the sheep\textquoteright s mass is greater than the threshold $f(N)$, a normalized force vector, $F^t_{\beta_jcd}$, is applied for the shepherd as a velocity vector in order to reach a collecting point. This point is positioned behind the outer or furthest sheep on the line drawn from the center of the sheep\textquoteright s mass to the furthest sheep.

\item \emph{Jittering behaviour} $\sigma_3$: To avoid an impasse during moving, a small random noise  $F^t_{\beta_j\epsilon}$ with weight $W_{e\beta_j}$, is added to the total force.

\end{itemize}

The total force $F^t_{\beta_j}$ of the shepherd $\beta_j$ (total force behaviour $\sigma_8$) is a weighted combination of the forces produced by the driving/collecting behaviour and the jittering behaviour.  This total force is shown in Equation~\ref{eq:ShepherdTotalForceEquation}

\begin{equation}\label{eq:ShepherdTotalForceEquation} F^t_{\beta_j} = F^t_{\beta_jcd} + W_{e\beta_j}  F^t_{\beta_j\epsilon} \end{equation}
}

\item {For sheep $\pi_i$}:

\begin{itemize}
\item \emph{Escaping behaviour} $\sigma_4$: This behaviour happens when the distance between the sheep $\pi_i$ at position $P^t_{\pi_i}$ and the shepherd $\beta_j$ at position $P^t_{\beta_j}$ is less than the sensing range, $R_{\pi\beta}$, a repulsive force $F^t_{\pi_i\beta_j}$ is provided the sheep $\pi_i$. The condition to trigger the behaviour is shown in Equation~\ref{eq:SheepDistanceToShepherdEquation}.

\begin{equation}\label{eq:SheepDistanceToShepherdEquation} \|P^t_{\pi_i}-P^t_{\beta_j}\| \le R_{\pi\beta} 
\end{equation}

\item \emph{Collision avoidance behaviour} $\sigma_5$: This behaviour happens when there is a repulsion between the sheep $\pi_i$ and the other sheep $\pi_{k\ne i}$. The condition of activating the repulsion force between the two sheep is that the distance between them is less than the sensing range among sheep, $R_{\pi\pi}$. This condition is shown in Equation~\ref{eq:SheepDistanceToOtherSheepEquation}.

\begin{equation}\label{eq:SheepDistanceToOtherSheepEquation} \exists {k}, \ such \ that \ \|P^t_{\pi_i}-P^t_{\pi_{k}}\| \le R_{\pi\pi} 
\end{equation}

Then, we have the summed force vectors, $F^t_{\pi_i\pi_{-i}}$, from all the other sheep within the threshold range, $R_{\pi\pi}$, applied onto sheep $\pi_i$.

\item \emph{Grouping behaviour} $\sigma_6$: This behaviour appears when the sheep $\pi_i$ under a force $F^t_{\pi_i\Lambda^t_{\pi_i}}$ will be attracted to move towards the center of the mass of its sheep neighbors, $\Lambda^t_{\pi_i}$.

\item \emph{Jittering behaviour} $\sigma_7$: Similar to the jittering behaviour of each shepherd,  to avoid impasse, a small random noise is added to the total force $F^t_{\pi_i\epsilon}$ with weight $W_{e\pi_i}$.

\end{itemize}

The total force, $F^t_{\pi_i}$, of the sheep $\pi_i$ is represented by a weighted sum of individual force vectors $F^t_{\pi_i\beta_j}$, $F^t_{\pi_i\pi_{-i}}$, 
$F^t_{\pi_i{\Lambda^t_{\pi_i}}}$, and $F^t_{\pi_i\epsilon}$; that is,

\begin{equation}\label{eq:SheepTotalForceEquation}
\begin{aligned}
F^t_{\pi_i}  = W_{\pi_\upsilon}  F^{t-1}_{\pi_i} +
W_{\pi \Lambda} F^t_{\pi_i{\Lambda^t_{\pi_i}}} + W_{\pi\beta} F^t_{\pi_i\beta_j}  + 
\\
W_{\pi\pi}  F^t_{\pi_i\pi_{-i}}  + W_{e\pi_i} F^t_{\pi_i\epsilon} 
\end{aligned}
\end{equation}

\end{enumerate}

The shepherds\textquoteright and sheep\textquoteright s positions are calculated according to Equations~\ref{eq:UpdatedShepherdPositionEquation}, and~\ref{eq:UpdatedSheepPositionEquation}. Meanwhile, the given $S^t_{\beta_j}$ and $S^t_{\pi_i}$ are the speed of the shepherd $\beta_j$ and the speed of the sheep $\pi_i$ at time step $t$. In the original Str\"{o}mbom approach, the speeds of both the shepherds and sheep are constant.\\
\begin{equation}\label{eq:UpdatedShepherdPositionEquation}
P^{t+1}_{\beta_j} = P^{t}_{\beta_j} + S^t_{\beta_j} F^t_{\beta_j}
\end{equation}
\begin{equation}\label{eq:UpdatedSheepPositionEquation}
P^{t+1}_{\pi_i} = P^t_{\pi_i} + S^t_{\pi_i} F^t_{\pi_i}
\end{equation}

\section{A Proposed Evaluation Framework.}\label{sect:framework}

In this paper, we evaluate the performance of the shepherd produced by the Str\"{o}mbom approach~\cite{strombom2014solving} under the two noises: the actuation ($\lambda$) and perception ($\alpha$). The actuation noise appears when the sheep move randomly around the location they are supposed to move to. The range of the random movement decides the degree of this noise. Meanwhile, the perception noise happens when there is deviation between the sheep\textquoteright s actual position and the position that the shepherd observes. The deviation range between these two positions defines the degree of the perception noise. 

The standard normal distribution, with a mean of zero and standard deviation of 1, is used to create the actuation and perception noises. The procedure of updating the position of sheep $\pi_i$ under the actuation noise, called $ActP^{t+1}_{\pi_i}$, is shown in Equation~\ref{eq:addActuationNoise}.

\begin{equation}\label{eq:addActuationNoise} 
ActP^{t+1}_{\pi_i} = P^t_{\pi_i} + S^t_{\pi_i}\times( F^t_{\pi_i} +  \lambda\times StandardNormal())
\end{equation}

For the perception noise, the perceived position of sheep $\pi_i$ at timestep $t+1$ is denoted $PerP^{t+1}_{\pi_i}$. This position is sensed by the shepherd by Equation~\ref{eq:addPerceptionNoise}.

\begin{equation}\label{eq:addPerceptionNoise} 
PerP^{t+1}_{\pi_i} = ActP^{t+1}_{\pi_i} +  \alpha\times StandardNormal()
\end{equation}  

According to the Str\"{o}mbom approach~\cite{strombom2014solving}, there are two main behaviours: collecting and driving. To switch between these behaviours, the shepherd needs to check whether any sheep is further from the center of mass than the threshold ($f(N)$) as calculated in Equation~\ref{eq:ShepherdBehaviorSelectionThresholdEquation}. We aim to identify appropriate thresholds ($f(N)$) of triggering between the collecting and driving behaviours, informed by the estimates of noise levels, in order to improve the performance of the shepherd.

To find these appropriate thresholds, we attempt values above and below the threshold of the Str\"{o}mbom approach shown in Equation~\ref{eq:ShepherdBehaviorSelectionThresholdEquation} by decreasing or increasing a constant, called $\Delta_{f}$. In our evaluation, we conduct three decreasing levels of this constant ($-1,-2,-3$) and similarly three increasing levels of this constant ($1,2,3$). We refer to the constant as a threshold due to its impact on Str\"{o}mbom\textquoteright s thresholding function to switch behaviour. These six levels will be multiplied with the $\Delta_f$. Thus, we have seven threshold values from $f(N) - 3\times\Delta_f$ to $f(N) + 3\times\Delta_f$ as shown in Table~\ref{tab:thresholdvalues}. It can be understood that when the threshold value increases, the collecting frequency will decrease, and the shepherd focuses on the driving behaviour. We set $\Delta_f=5(meter)$ in this paper.\\ 

\begin{table}[ht]
\centering
\caption{The Threshold to Switch Between Collecting and Driving.} \label{tab:thresholdvalues}
\begin{tabular}{c|c|c}
Collection Frequency & Parameter     &  Value (meter) \\ \hline        
Extreme & $f_{-3}$ &  $f(N) - 3\times\Delta_f$\\
Very High & $f_{-2}$ &  $f(N) - 2\times\Delta_f$\\
High & $f_{-1}$ &  $f(N) - 1\times\Delta_f$\\
Normal & $f_{0}$ &  $f(N)$\\
Infrequent & $f_{+1}$ &  $f(N) + 1\times\Delta_f$\\
Very Infrequent & $f_{+2}$ &  $f(N) + 2\times\Delta_f$\\
Rare & $f_{+3}$ &  $f(N) + 3\times\Delta_f$\\
\hline
\end{tabular} 
\end{table}

For both actuation and perception noises, we set six noise levels increasing $0.01$ from $0.01$ to $0.06$ and $0.1$ from $0.1$ to $0.6$, respectively. To add these noises to the operation, we multiply these noise levels with a fixed change value, $\Delta_{n}$, which is set to the investigated maximum threshold value $f_{+3}$. As well as the six noise levels for the perception noise, we also investigate the performance of the shepherd in the same perception condition of the Str\"{o}mbom model~\cite{strombom2014solving} without noise. Hence, we have seven noise levels for both as given in Table~\ref{tab:noisealphalevels} and \ref{tab:noiseactuationlevels}.

\begin{table}[ht]
\centering
\caption{Levels of Perception Noise ($\alpha$)} \label{tab:noisealphalevels}
\begin{tabular}{c|c|c}
Level & Perception Noise    & Value (meter)         \\ \hline Noise Free & $\alpha_0$  & $0$\\
Very little &  $\alpha_1$  & $0.1\times\Delta_n$\\
Little & $\alpha_2$ & $0.2\times\Delta_n$\\
Small & $\alpha_3$  & $0.3\times\Delta_n$\\
Medium & $\alpha_4$  & $0.4\times\Delta_n$\\
High & $\alpha_5$  & $0.5\times\Delta_n$\\
Very High & $\alpha_6$  & $0.6\times\Delta_n$\\
\end{tabular} 
\end{table}

\begin{table}[ht]
\centering
\caption{Levels of Actuation Noise ($\lambda$).} \label{tab:noiseactuationlevels}
\begin{tabular}{c|c|c|c}
Level & Actuation Noise & Value (meter)         \\ \hline Noise Free & $\lambda_0$ & $0$\\
Very little &  $\lambda_1$ & $0.01\times\Delta_n$\\
Little &  $\lambda_2$ & $0.02\times\Delta_n$\\
Small &$\lambda_3$ & $0.03\times\Delta_n$\\
Medium &  $\lambda_4$ & $0.04\times\Delta_n$\\
High & $\lambda_5$ & $0.05\times\Delta_n$\\
Very High & $\lambda_6$ & $0.06\times\Delta_n$\\
\end{tabular} 
\end{table}

In this work, the perception noise values are considerably higher than that of the actuation noise. The reason is because under  actuation noise, the sheep also have repulsion and attraction forces among them; thus, the movement of the sheep under the actuation noise is more spread. Meanwhile under the perception noise, the shepherd has a large sensing range (65 meters in the Str\"{o}mbom approach), and then it is sill able to control the sheep acceptably without reaching the true collecting and driving points. Therefore, the perception noise values need to be larger in order to assess its impact on performance.

The performance ($PF$) of the shepherd is validated based on combining the three above factors: the changing radius of the mass ($f$), and the two noise conditions ($\lambda$ and $\alpha$). This relation is illustrated in Equation~\ref{eq:relationthreefactors} in which the function - $g$ includes three variables ($f,\lambda, \alpha$).  

\begin{equation}\label{eq:relationthreefactors}
PF_{\beta_i} = g(f,\lambda,\alpha) 
\end{equation}
with $\beta_i$ is the shepherd $i$-th.
In the next section, we provide the design of the experiments in this paper.

\section{Experiments}\label{sect:experiment}

In this paper, we simulate the environment given by the Str\"{o}mbom model~\cite{strombom2014solving} as introduced in Section~\ref{sect:shepherding}. The same parameters regarding the environment initialization and the interaction between agents for the simulation are listed in Table~\ref{tab:EnvironmentalParameters}. 

\begin{figure}[!ht]
    \centering
    \includegraphics[width=0.3\textwidth]{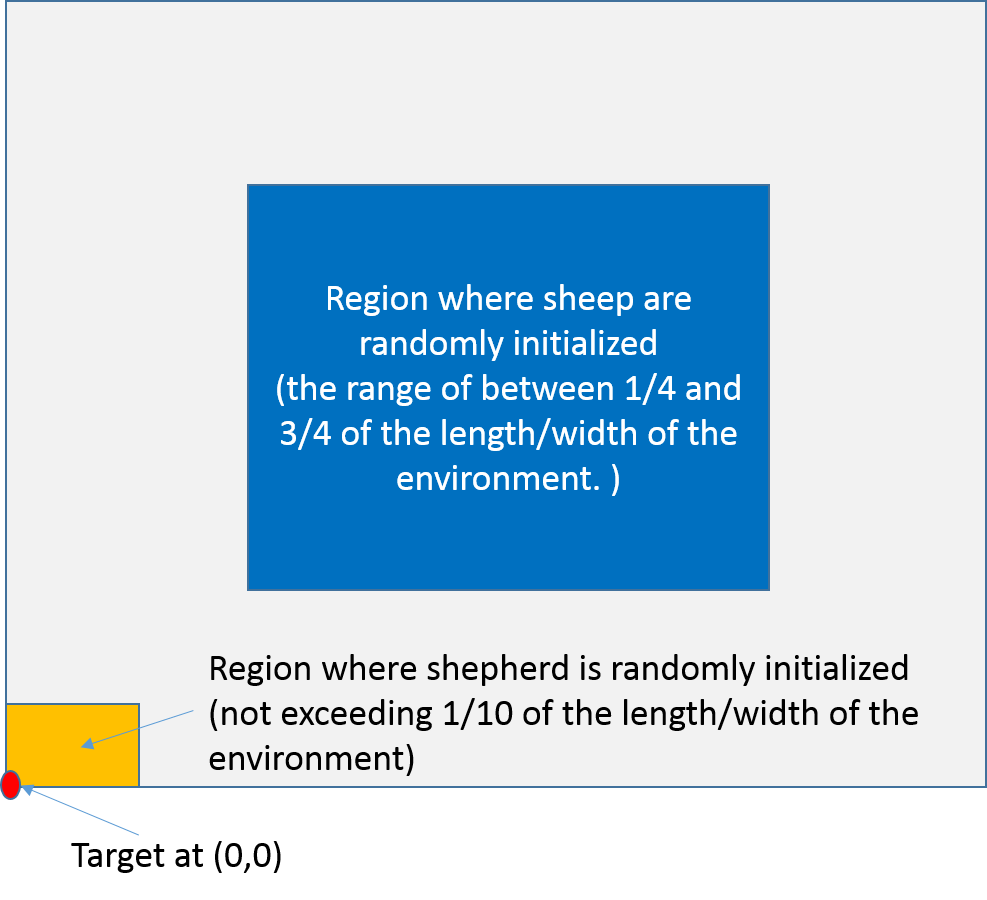}
    \caption{The experiment environment.}
    \label{fig:envexperimnet}
\end{figure}

In each simulation, a given number of sheep are randomly initialized at the centre of the paddock with their coordinates ranging between 1/4 and 3/4 of the length/width of the environment. The shepherd is randomly initialised at the lower left corner, with their coordinates not exceeding 1/10 of the length/width of the environment, near the target position (at $(0,0)$). The shepherd\textquoteright s mission is to collect outer sheep into a group and herd the entire sheep towards the target. The mission is achieved when all of the sheep have reached the target within a limit of 1000 steps, and the simulation ends. The illustration of the environment is shown in Figure~\ref{fig:envexperimnet}. 

\begin{table}
\caption{Environmental parameters in the simulation.}
\begin{tabular}{c|p{1.8in}|c}
Parameter                     & \multicolumn{1}{c|}{Meaning}                                                                   & Value          \\ \hline
$L$ & Length and Width of Environment & 150\\
$N$& Number of Sheep & 100 \\
$M$& Number of Shepherds & 1 \\
$R_{\pi\beta}$& Sensing range of a sheep for the shepherd & 65\\
$R_{\pi\pi}$ & Sensing range of a sheep for another sheep & 2 \\
$W_{\pi\pi}$ & Sheep repulsion strength from other sheep & 2\\
$W_{\pi\beta}$ & Sheep repulsion strength from the shepherd & 1              \\
$W_{\pi \Lambda}$  & Sheep attraction strength to sheep centre of mass  & 1.05  \\
$W_{\pi_\upsilon}$  & Inertial strength of sheep previous direction  & 0.5 \\
$W_{e\pi_i}$ & Strength of sheep movement noise  & 0.3 \\
$W_{e\beta_j}$ & Strength of the shepherd movement noise  & 0.3  \\
$\vert\Omega_{\pi_i\pi}\vert$ & Number of sheep (neighborhood) a sheep can sense   & 25 \\
$S_\pi$ & Maximum speed of sheep & 1 \\
$S_{\beta}$   & Maximum speed of the shepherd & 2\\
$\mathbb{D}$  & Minimum distance between the sheep\textquoteright s global centre of mass and the target for successful mission & 5             
\end{tabular}
\label{tab:EnvironmentalParameters}
\end{table}

\subsection{Experimental Setups}

In this paper, we evaluate the performance of the shepherd of the Str\"{o}mbom model~\cite{strombom2014solving} under the two noises: actuation - $\lambda$ and perception -$\alpha$. Furthermore, we aim to identify the appropriate thresholds ($f$) of triggering between the collecting and driving behaviours that might lead to higher performance for the shepherd under these two noises. Thus, in total we conduct $7\times7\times7 = 343$  setups in which there are 7 changing levels of the threshold ($f$), 7 perception noise levels ($\alpha$), and 7 actuation noise levels ($\lambda$). 

\subsection{Evaluation Metrics}\label{sect:EvaluationMetrics}

In order to evaluate the performance of the shepherd under these three factors ($f,\lambda, \alpha$), we use two assessment metrics as below:

\begin{itemize}
    \item \textbf{Number of steps (NS)}: the number of time steps for the sheep to be herded to the target location. 
    \item \textbf{Success rate ($\%$) (SR)} is the percentage of missions completed computed on a number of testing cases. The mission success is achieved when all the sheep are collected and driven to the goal position.
    \item \textbf{Standard Error of Mean (SEM)} of NS indicates stability of the evaluation. This metric is calculated in Equation~\ref{eq:SEM}.
    \begin{equation}\label{eq:SEM}
    SEM = Std/\sqrt{n} 
    \end{equation}
    where $Std$ is the standard deviation of the number of steps, and $n$ is the number of episodes.
    \item \textbf{Standard Error Percentage (SEM-P)} of Mean indicates which episode the evaluation should be stopped when the SEM-P is less than a small threshold (in this work, we choose 3$\%$) of the mean.
\end{itemize}

\section{Results and Discussion}\label{sect:results}

In this paper, we conduct 300 random testing episodes for each of the 343 setups. This number of testing episodes enables our evaluations to be able to reach a level of stability to ensure appropriateness of the analyses and precision in assessing the performance of the shepherd. The evaluation achieves our stability criterion when the SEM-P value of the performance of the shepherd in setups is less than the threshold of 3$\%$ of the mean.

\begin{figure*}[!ht]
    \centering
 \subfigure[SEM of $f$ at $\lambda_0$ and $\alpha_0$]
    {
        \includegraphics[width=0.25\textwidth]{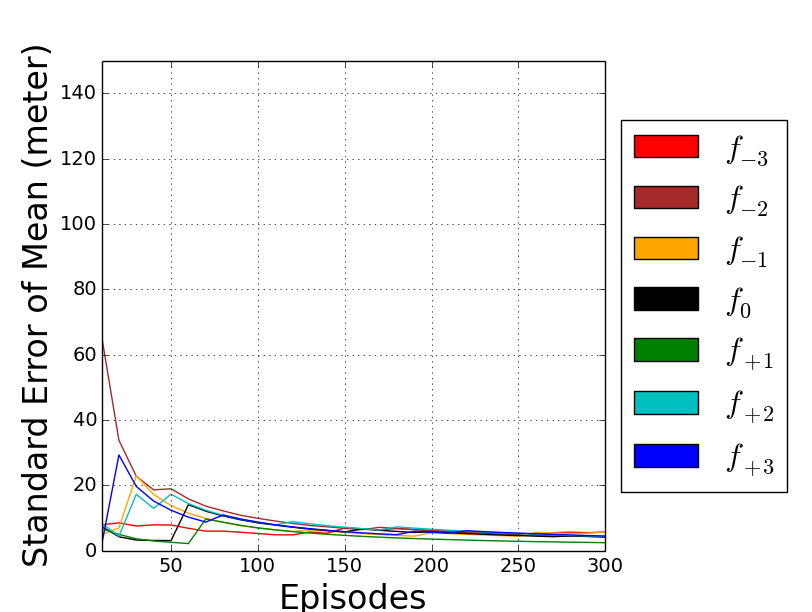}
        \label{fig:driving4x4_sub}
    }%
     \subfigure[SEM of $f$ at $\lambda_0$ and $\alpha_1$]
    {
        \includegraphics[width=0.25\textwidth]{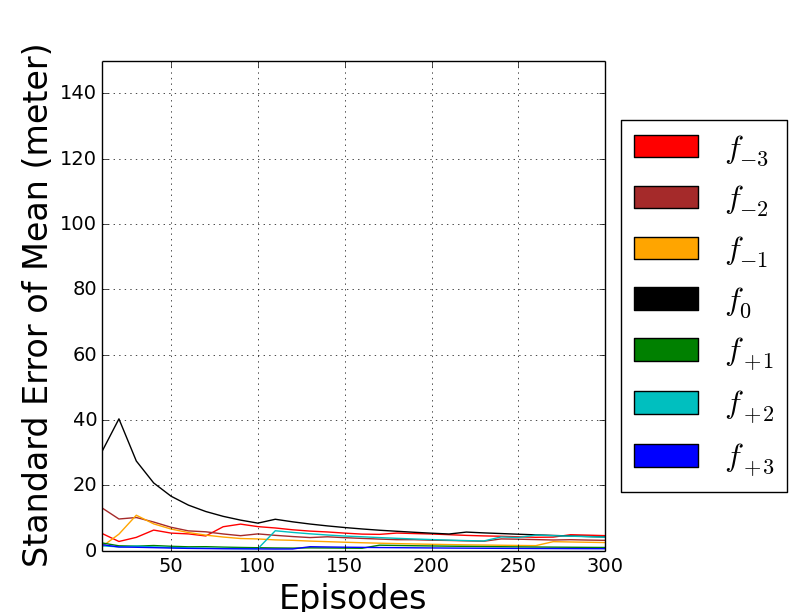}
        \label{fig:driving4x4_sub}
    }%
     \subfigure[SEM of $f$ at $\lambda_0$ and $\alpha_2$]
    {
        \includegraphics[width=0.25\textwidth]{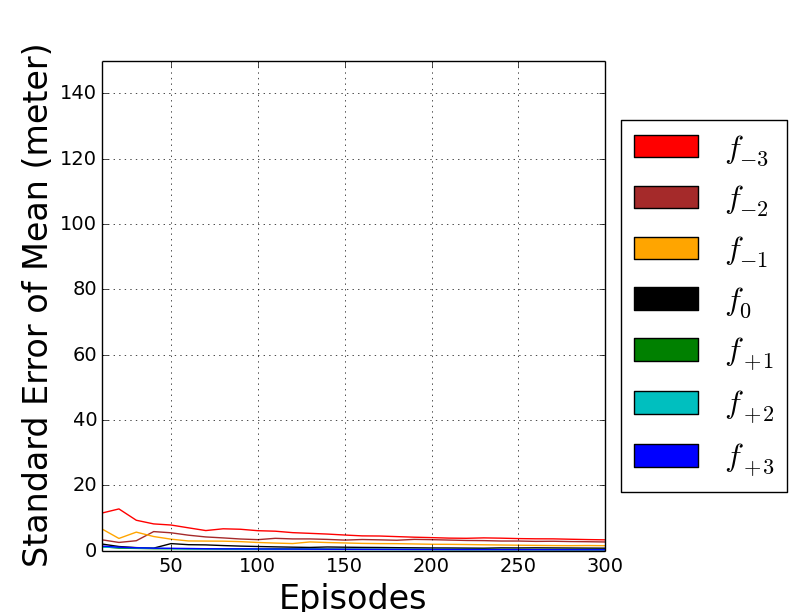}
        \label{fig:driving4x4_sub}
    }%
    \subfigure[SEM of $f$ at $\lambda_0$ and $\alpha_3$]
    {
        \includegraphics[width=0.25\textwidth]{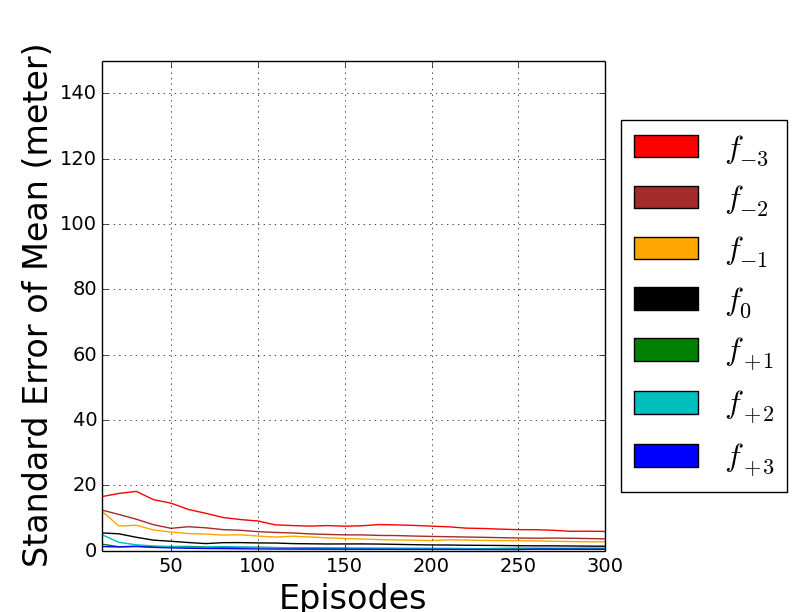}
        \label{fig:driving4x4_sub}
    }%
    \\
     \subfigure[SEM of $f$ at $\lambda_0$ and $\alpha_4$]
    {
        \includegraphics[width=0.25\textwidth]{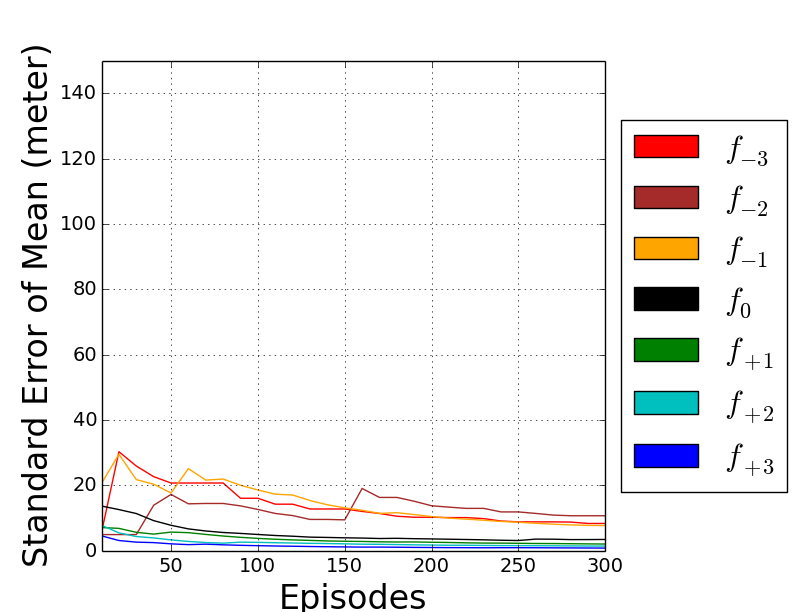}
        \label{fig:driving4x4_sub}
    }%
    \subfigure[SEM of $f$ at $\lambda_0$ and $\alpha_5$]
    {
        \includegraphics[width=0.25\textwidth]{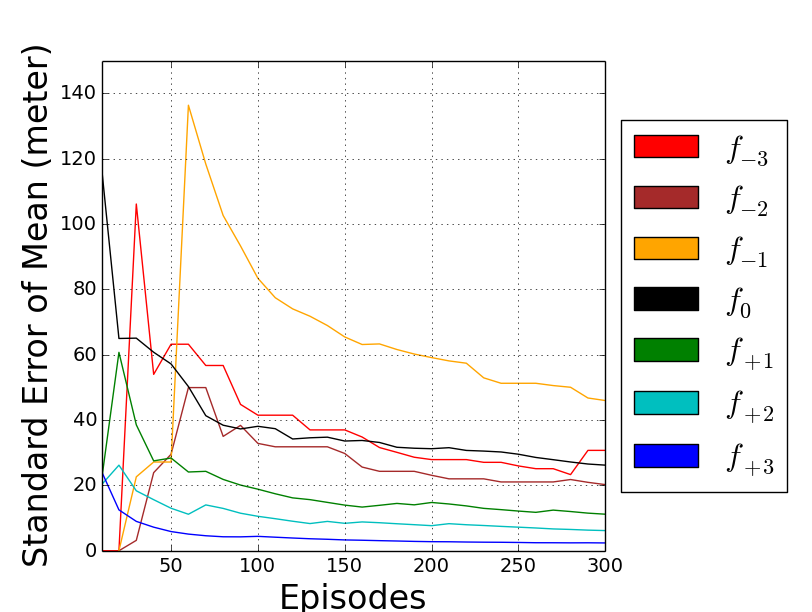}
        \label{fig:driving4x4_sub}
    }%
    \subfigure[SEM of $f$ at $\lambda_0$ and $\alpha_6$]
    {
        \includegraphics[width=0.25\textwidth]{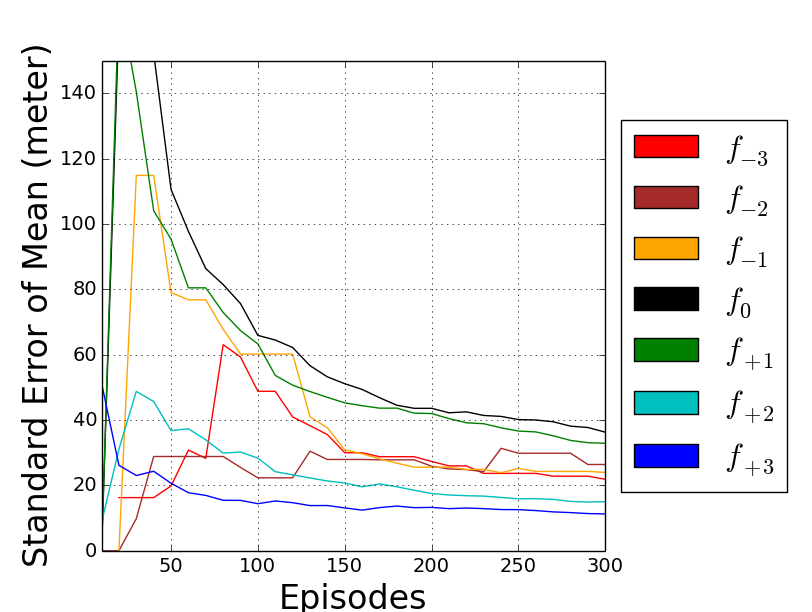}
        \label{fig:driving4x4_sub}
    }%
     \caption{Standard Error of Mean to Evaluate Stability of Alpha ($\alpha$) with Different Thresholds or Collecting Frequency ($f$) at Lambda ($\lambda_0$) in 300 Episodes.}
    \label{fig:alphax-lamda0}
\end{figure*}

\begin{figure}[!ht]
    \centering
    \subfigure[SEM-P of Mean of $f$ at ($\lambda_0$, $\alpha_5$)]
    {
        \includegraphics[width=0.245\textwidth]{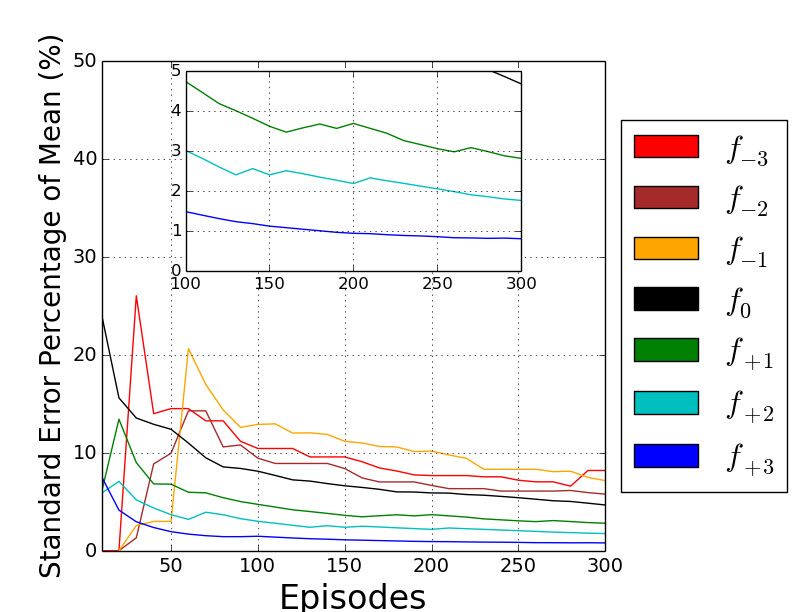}
        
    }%
    \subfigure[SEM-P of $f$ at ($\lambda_0$, $\alpha_6$)]
    {
        \includegraphics[width=0.245\textwidth]{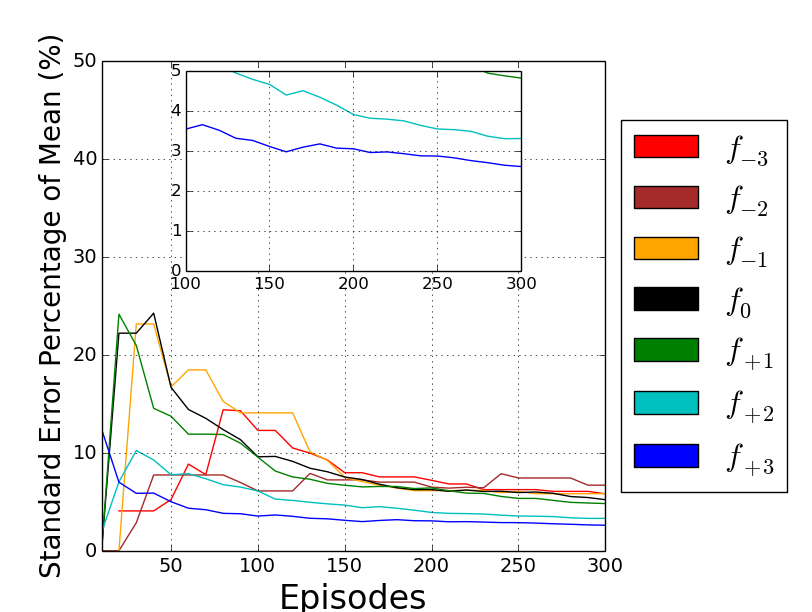}
        
    }%
     \caption{Standard Error of Mean to Evaluate Stability of Alpha ($\alpha$) with Different Thresholds or Collecting Frequency ($f$) at Lambda ($\lambda_0$) in 300 Episodes.}
    \label{fig:alphax-lamda0-semp}
\end{figure}

Firstly, we investigate the effects of the actuation ($\lambda$) and perception ($\alpha$) noises on the performance of the shepherd. We show the evidences of the stability of the evaluation when the values of the SEM and SEM-P of the setups decrease gradually and maintain stability by the end of the 300 episodes. According to the SEM values of $f$, Figure~\ref{fig:alphax-lamda0} shows that when there is no actuation noise ($\lambda_0$), the performance of the shepherd is stably evaluated and maintained after 150 episodes at pairs of $\lambda_0$ and the perception noise levels $\alpha$ lower than the high level. At the high ($\alpha_5$) and very high ($\alpha_6$) perception noise levels, the stability as measured by the performance of the shepherd is not obtained, and then, the reliability of our ability to estimate the performance might be imprecise. The performance instability at the high and very high levels of the perception noise can be seen in Figure~\ref{fig:alphax-lamda0-semp}. In this Figure, we can see that just only two setups of the very infrequent $f_{+2}$ and the rare of collecting frequency $f_{+3}$ at the high level ($\alpha_5$), and the setup of the rare of collecting frequency $f_{+3}$ at the very high level ($\alpha_6$) have  SEM-P values below the stop point of 3$\%$.

\begin{figure*}[!ht]
    \centering
 \subfigure[SEM of $f$ at $\lambda_0$ and $\alpha_0$]
    {
        \includegraphics[width=0.25\textwidth]{figures/lambda0-alpha0-sem.png}
        \label{fig:driving4x4_sub}
    }%
     \subfigure[SEM of $f$ at $\lambda_1$ and $\alpha_0$]
    {
        \includegraphics[width=0.25\textwidth]{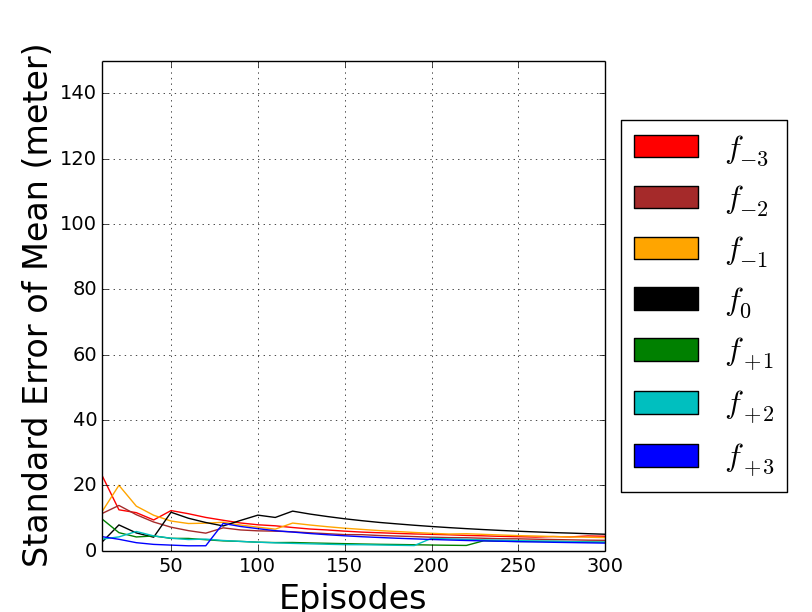}
        \label{fig:driving4x4_sub}
    }%
     \subfigure[SEM of $f$ at $\lambda_2$ and $\alpha_0$]
    {
        \includegraphics[width=0.25\textwidth]{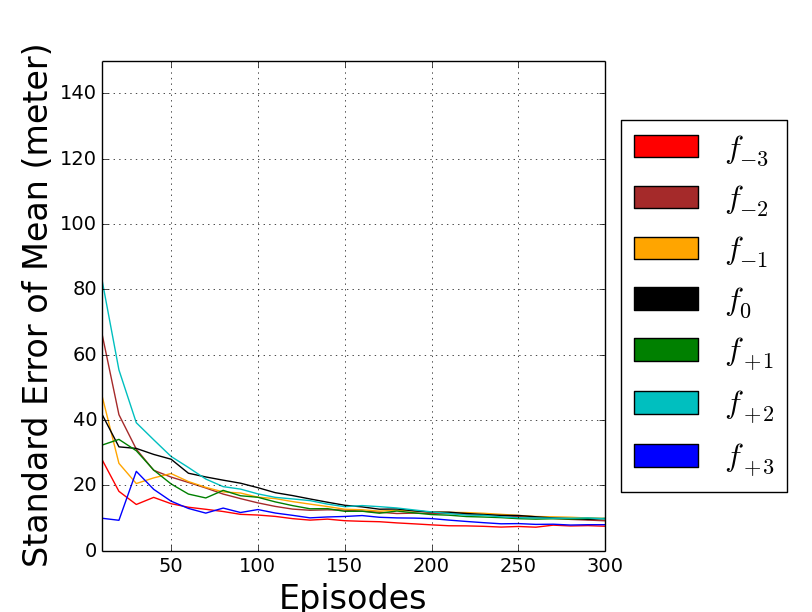}
        \label{fig:driving4x4_sub}
    }%
    \\
    \subfigure[SEM of $f$ at $\lambda_3$ and $\alpha_0$]
    {
        \includegraphics[width=0.25\textwidth]{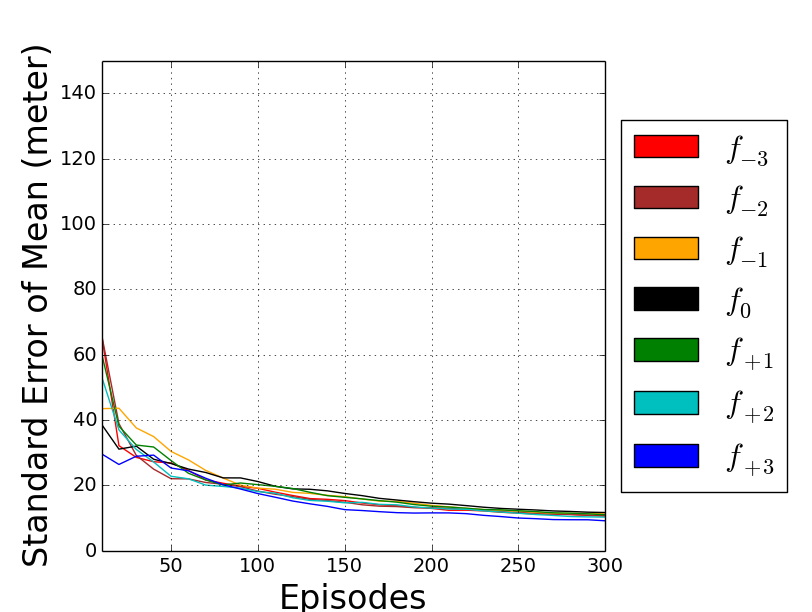}
        \label{fig:driving4x4_sub}
    }%
     \subfigure[SEM of $f$ at $\lambda_4$ and $\alpha_0$]
    {
        \includegraphics[width=0.25\textwidth]{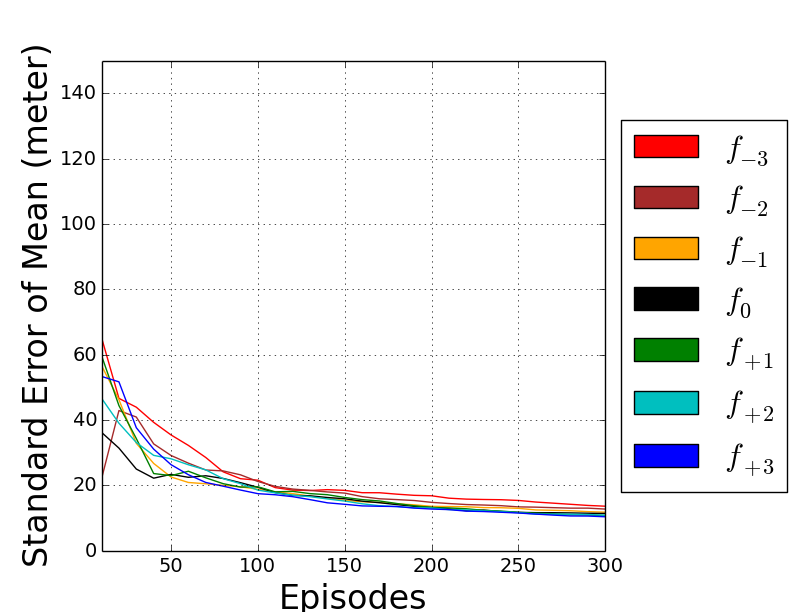}
        \label{fig:driving4x4_sub}
    }%
    \subfigure[SEM of $f$ at $\lambda_5$ and $\alpha_0$]
    {
        \includegraphics[width=0.25\textwidth]{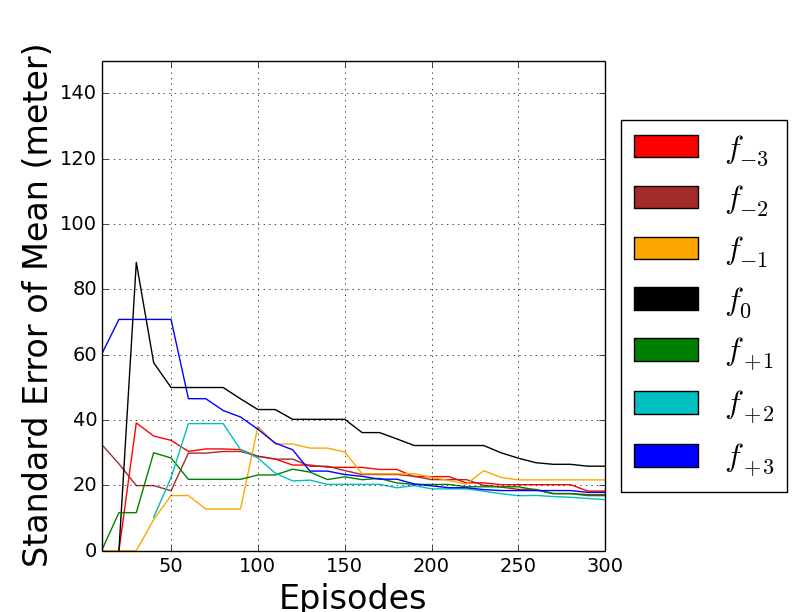}
        \label{fig:driving4x4_sub}
    }%
     \caption{Standard Error of Mean to Evaluate Stability of Lambda ($\lambda$) with Different Thresholds or Collecting Frequency ($f$) at Alpha ($\alpha_0$) in 300 Episodes.}
    \label{fig:alpha0-lamdax}
\end{figure*}

\begin{figure}[!ht]
    \centering
     \subfigure[SEM-P of Mean of $f$ at $\lambda_4$ and $\alpha_0$]
    {
        \includegraphics[width=0.245\textwidth]{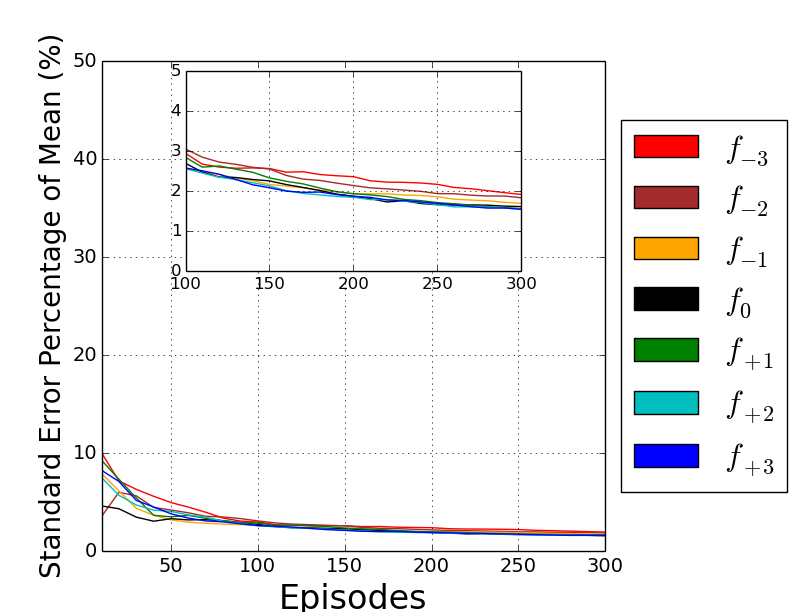}
        \label{fig:driving4x4_sub}
    }%
     \subfigure[SEM-P of Mean of $f$ at $\lambda_5$ and $\alpha_0$]
    {
        \includegraphics[width=0.245\textwidth]{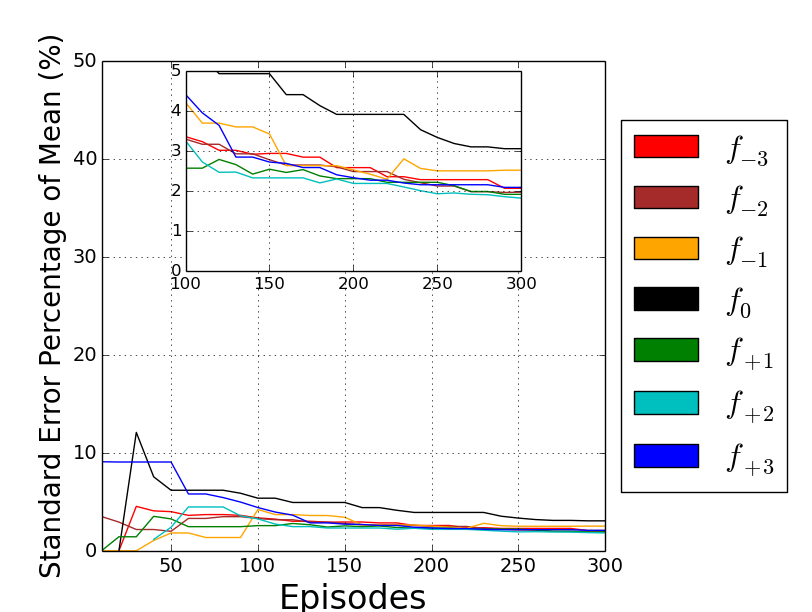}
        \label{fig:driving4x4_sub}
    }%
     \caption{Standard Error of Mean to Evaluate Stability of Lambda ($\lambda$) with Different Thresholds or Collecting Frequency ($f$) at Alpha ($\alpha_0$) in 300 Episodes.}
    \label{fig:alpha0-lamdax-semp}
\end{figure}

Similarly, according to the SEM values of $f$, Figure~\ref{fig:alpha0-lamdax} shows when there is no perception noise, the shepherd exhibits stable behaviour over 150 episodes  at pairs of $\lambda$ from the noise free level to the medium level and $\alpha_0$. When high actuation noise ($\lambda_5$) is applied, the behaviour of the shepherd fluctuates more in the first 150 episodes, and takes another 150 episodes to reach to the stable point wherein all the SEM-P values of the threshold $f$ reduces to below 3$\%$ can be seen in Figure~\ref{fig:alpha0-lamdax-semp}. Furthermore, with the very high actuation noise ($\lambda_6$), the shepherd is not able to achieve the mission.

\begin{figure}[!ht]
    \centering
    \subfigure[SEM of $f$ at ($\lambda_2$, $\alpha_5$)]
    {
        \includegraphics[width=0.245\textwidth]{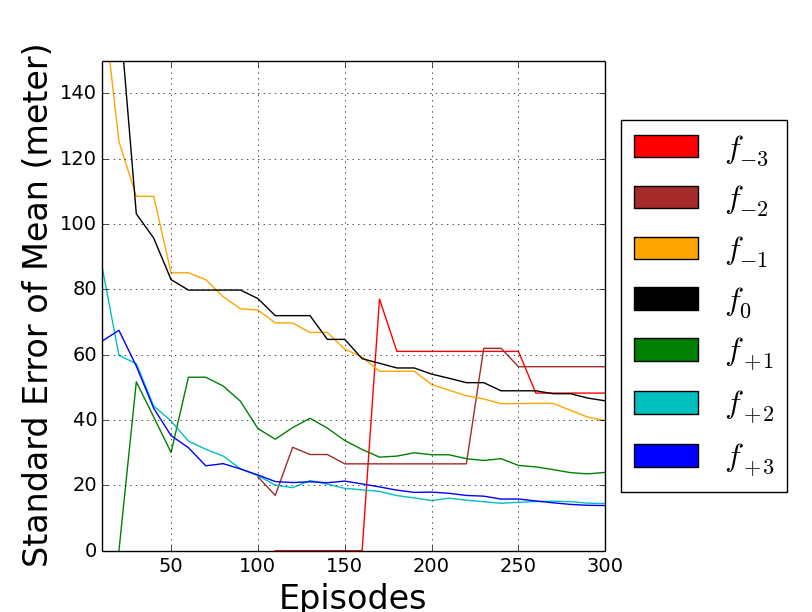}
        \label{fig:driving4x4_sub}
    }%
    \subfigure[SEM-P of $f$ at ($\lambda_2$, $\alpha_5$)]
    {
        \includegraphics[width=0.245\textwidth]{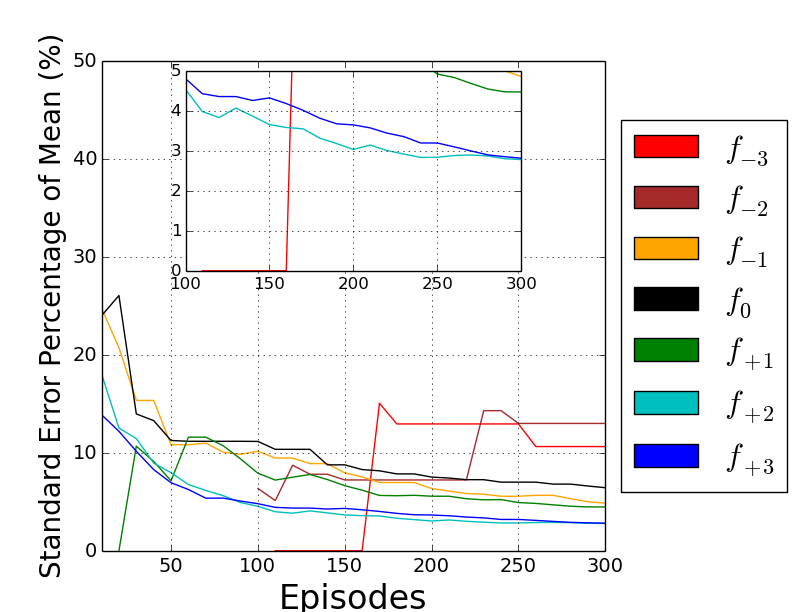}
        \label{fig:driving4x4_sub}
    }%
     \caption{Standard Error of Mean to Evaluate Stability of Alpha ($\alpha$) with Different Thresholds or Collecting Frequency ($f$) at Lambda ($\lambda_2$) in 300 Episodes.}
    \label{fig:alpha-lamdax-notworking-evaluation}
\end{figure}

Similar results coming from the other pairs of $\lambda$ and $\alpha$ show that the stability measured from performance of the shepherd is not reached at noise levels higher than medium in both actuation and perception. Especially, the shepherding task collapses at a very high level of actuation noise $\lambda_6$. Figure~\ref{fig:alpha-lamdax-notworking-evaluation} illustrates this instability when the majority of the SEM and SEM-P values show instability by the end of our evaluation and can not reduce to a stable point wherein the SEM-P values are below $3\%$. We note that in the unstable condition of the evaluation, the reliability to estimate the performance of the shepherd might be imprecise so in the next parts, we just focus on analysing the performance on the setups reaching stability.

\begin{figure}[!ht]
    \centering
    \subfigure[NS of f at $\alpha_0$ in 300 Episodes]
    {
        \includegraphics[width=0.25\textwidth]{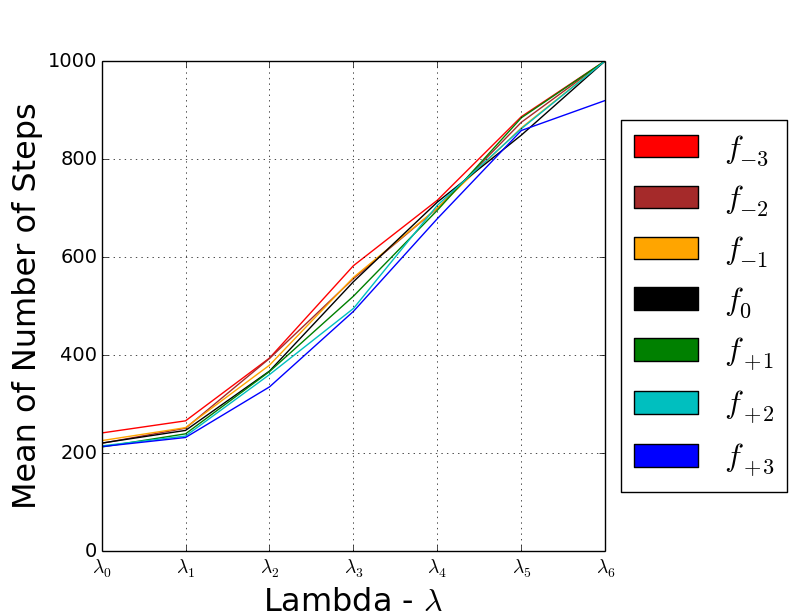}
        \label{fig:driving4x4_sub}
    }%
    \subfigure[NS of f at $\lambda_0$ in 300 Episodes]
    {
        \includegraphics[width=0.25\textwidth]{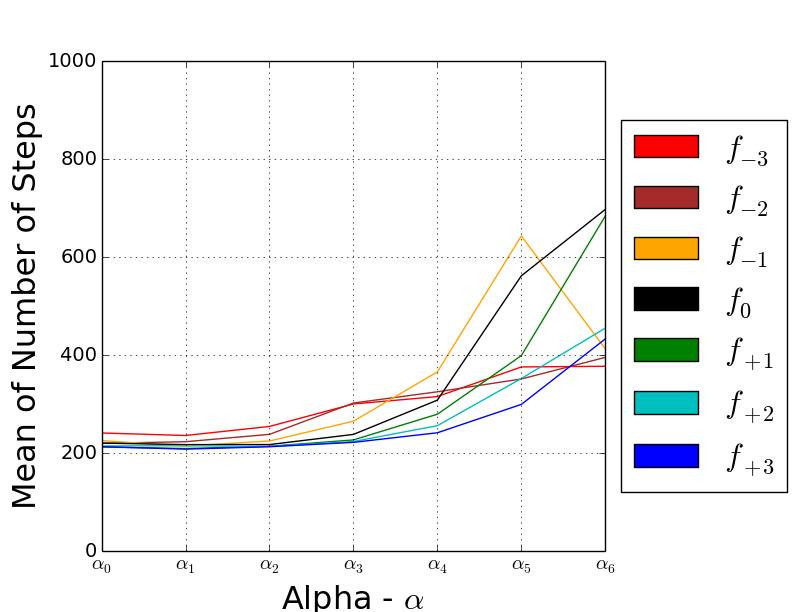}
        \label{fig:collecting4x4_sub}
    }%
    \\
    \subfigure[SR of f at $\alpha_0$ in 300 Episodes]
    {
        \includegraphics[width=0.25\textwidth]{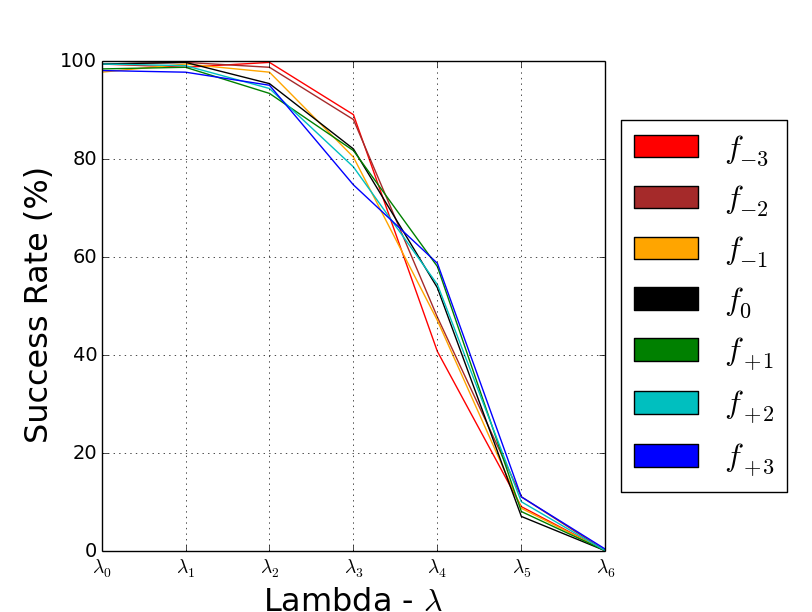}
        \label{fig:driving4x4_sub}
    }%
    \subfigure[SR of f at $\lambda_0$ in 300 Episodes]
    {
        \includegraphics[width=0.25\textwidth]{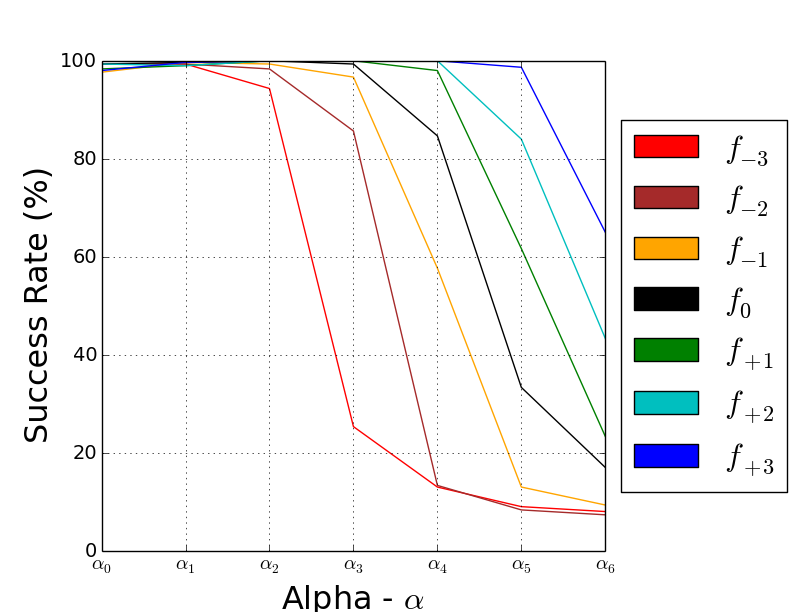}
        \label{fig:collecting4x4_sub}
    }%
    \caption{The Relationship between Lamda ($\lambda$), Alpha ($\alpha$), and the Threshold or the Collecting Frequency ($f$) in 300 Episodes when the Standard Error Percentage of Mean is below 3 percent (\%)}
    \label{fig:lamda0-alpha0}
\end{figure}

After validating the stability of the setups, we investigate how the actuation and perception noises impact the performance of the shepherd. Figure~\ref{fig:lamda0-alpha0} shows actuation noise ($\lambda$) impacting the performance of the shepherd more dramatically than the perception noise ($\alpha$). We can see that when under the noise-free condition of  perception $\alpha_0$, the shepherding task collapses at $\lambda_5$; meanwhile, without the actuation noise ($\lambda_0$), the shepherd is still able to successfully achieve the shepherding task until the very high level-$\alpha_7$ is reached with approximately 600 steps. It notes that the value of the actuation noise is considerably smaller than that of the perception noise (10 times). Furthermore, it is interesting that the change of the threshold $f$ does not impact drastically on the performance of the shepherd under the actuation noise; meanwhile, for the perception noise, this change has obvious effects on the shepherd\textquoteright s performance. With decreasing collecting frequency, the success rate is maintained at nearly $100\%$ and a smaller number of steps are required even though the perception noise increases. 

\begin{figure*}[!ht]
    \centering
    \subfigure[SR of $\alpha$ at $\lambda_1$]
    {
        \includegraphics[width=0.25\textwidth]{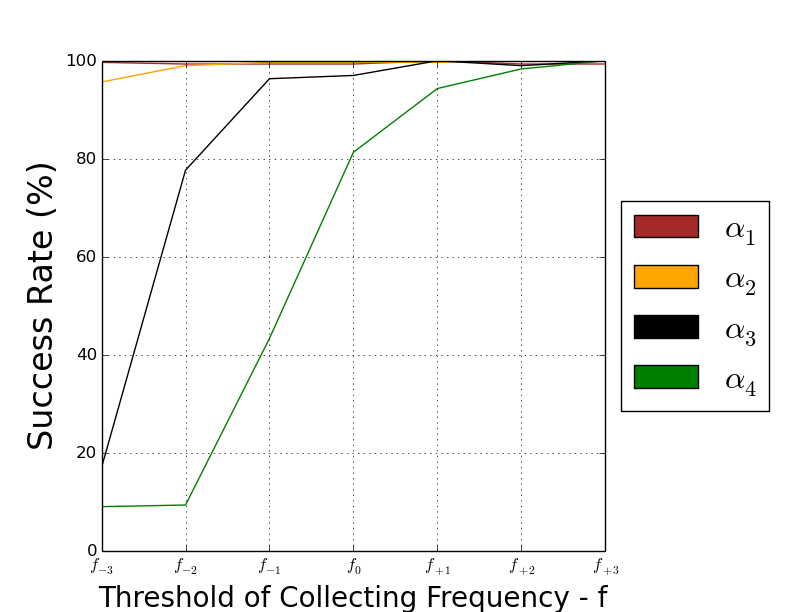}
        \label{fig:alpha-lambda1-fx-sr}
    }%
    \subfigure[SR of $\alpha$ at $\lambda_2$]
    {
        \includegraphics[width=0.25\textwidth]{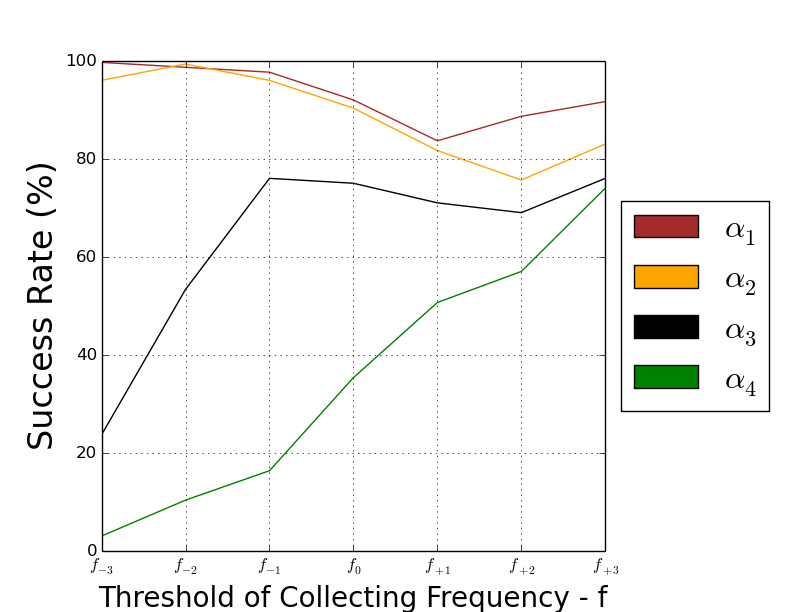}
        \label{fig:alpha-lambda2-fx-sr}
    }%
    \subfigure[SR of $\alpha$ at $\lambda_3$]
    {
        \includegraphics[width=0.25\textwidth]{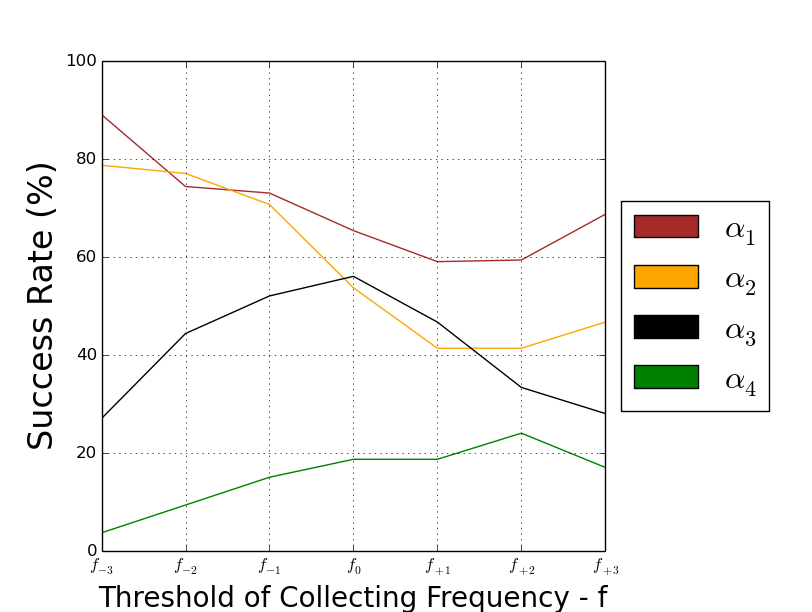}
        \label{fig:alpha-lambda3-fx-sr}
    }%
    \subfigure[SR of $\alpha$ at $\lambda_4$ ]
    {
        \includegraphics[width=0.25\textwidth]{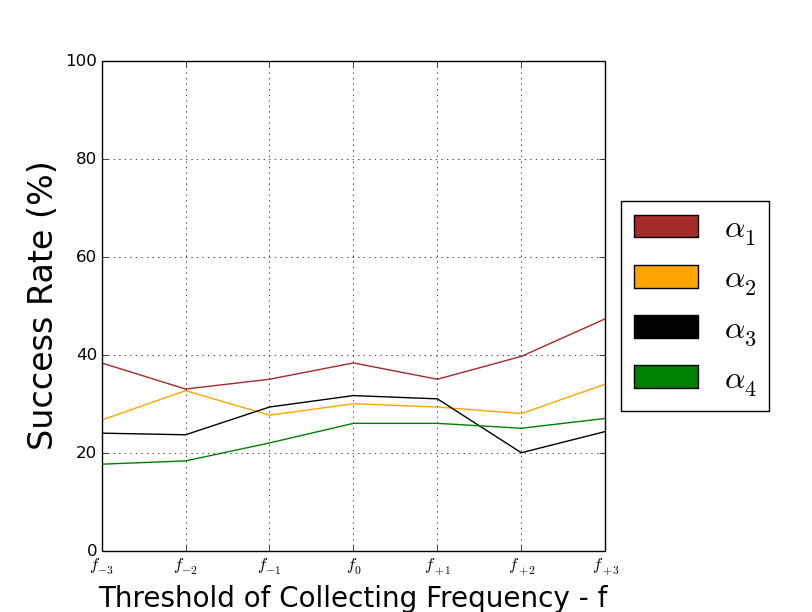}
        \label{fig:alpha-lambda4-fx-sr}
    }%
    \\
    \subfigure[NS of $\alpha$ at $\lambda_1$]
    {
        \includegraphics[width=0.25\textwidth]{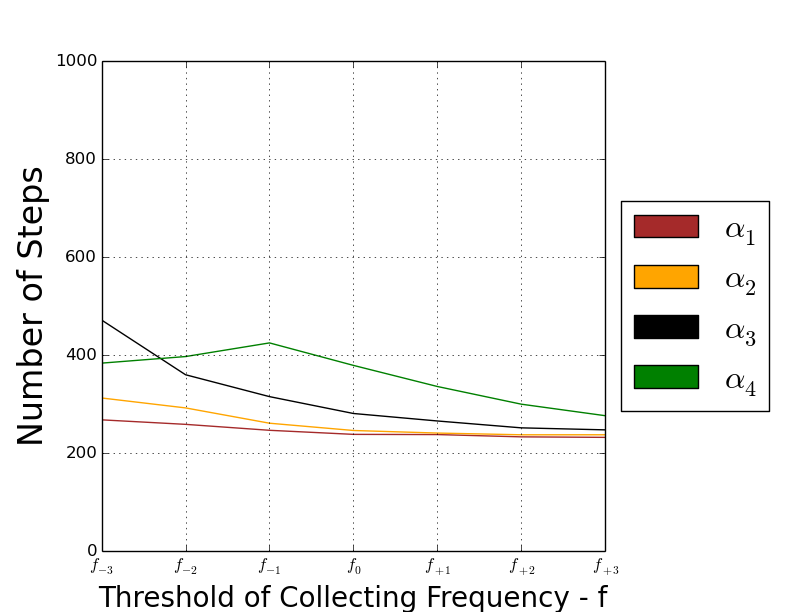}
        \label{fig:alpha-lambda1-fx-ns}
    }%
    \subfigure[NS of $\alpha$ at $\lambda_2$]
    {
        \includegraphics[width=0.25\textwidth]{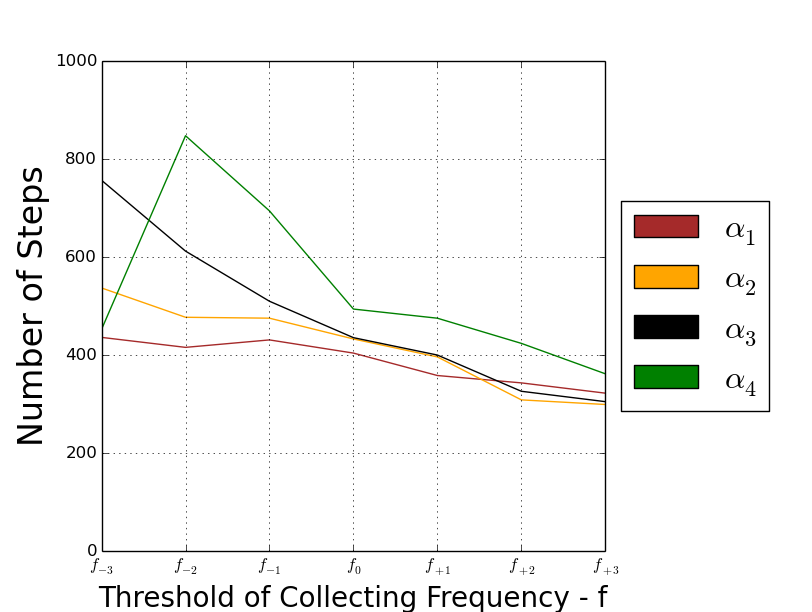}
        \label{ffig:alpha-lambda2-fx-ns}
    }%
    \subfigure[NS of $\alpha$ at $\lambda_3$]
    {
        \includegraphics[width=0.25\textwidth]{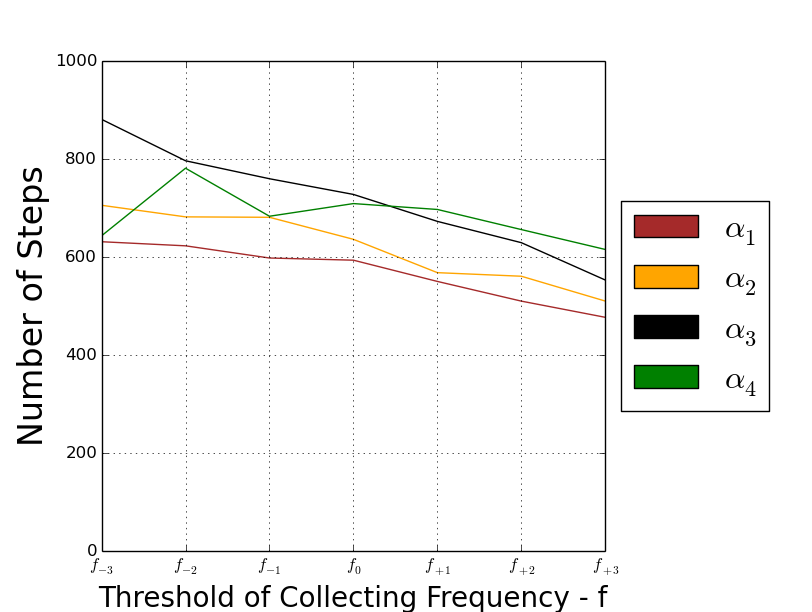}
        \label{fig:alpha-lambda3-fx-ns}
    }%
    \subfigure[NS of $\alpha$ at $\lambda_4$ ]
    {
        \includegraphics[width=0.25\textwidth]{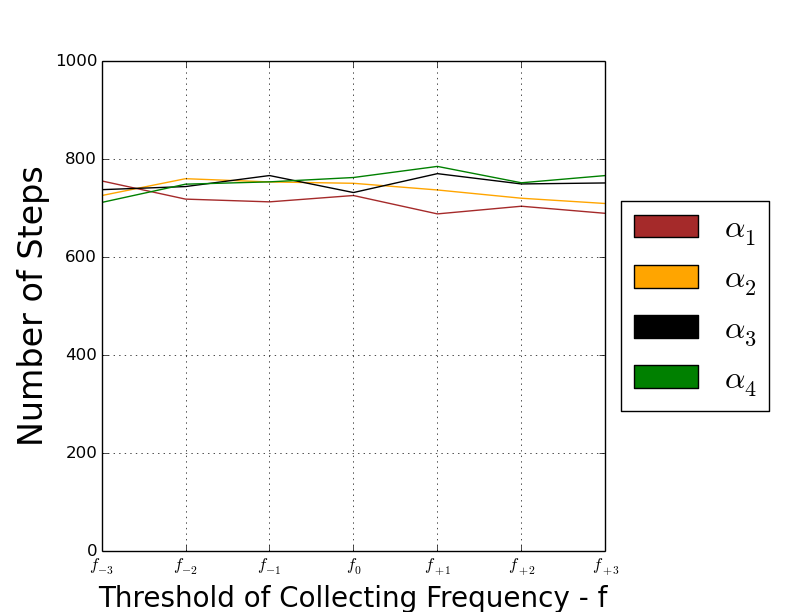}
        \label{fig:alpha-lambda4-fx-ns}
    }%
    \caption{The Effects of Lamda ($\lambda$) and Alpha ($\alpha$) on Different Thresholds or Collecting Frequency ($f$) in 300 Episodes when the Standard Error Percentage of Mean is below 3 percent (\%).}
    \label{fig:alpha-lamdax}
\end{figure*}

Besides evaluating the effect of the two types of noises on the performance of the shepherd, we conduct an additional evaluation between the changes of the collecting frequency and the performance. From this evaluation, a set of the appropriate thresholds $f$ leading to the shepherd\textquoteright s higher performance is provided in this paper. We focus on those setups achieving a stable performance. These setups vary in noise from very little noise ($\lambda_1$ and $\alpha_1$) to the medium noise ($\lambda_4$ and $\alpha_4$). Figure~\ref{fig:alpha-lamdax} shows the evidences of obtaining these appropriate thresholds. It can be seen that for the very little noise level of the actuation as shown in Figure~\ref{fig:alpha-lambda1-fx-sr}, the decrease in collecting frequency leads to considerably higher performance as well as the smaller number of steps when the perception noise increases. However, when the level of the actuation noise increases, under the little noise level of the perception as illustrated in Figure~\ref{fig:alpha-lambda2-fx-sr}, the shepherd should prefer the extreme and very high collecting frequency in order to have a higher success rate of nearly $100\%$ compared to approximately $95\%$ of the Str\"{o}mbom approach even though it takes more steps. Under the higher noise level of perception, the decreasing collecting frequency will improve performance in terms of both the success rate and the number of steps. 

Similarly, Figure~\ref{fig:alpha-lambda3-fx-sr} shows the case of having the small noise level ($\lambda_3$) of the actuation, the trend of choosing the small threshold $f$ under the very little or little perception noises allows an improvement in performance when comparing it with the Str\"{o}mbom approach. Furthermore, at a low perception noise level, it seems that the Str\"{o}mbom approach produces the best performance. Additionally, when the actuation noise is at a medium level, there is not considerable difference between the performance under the increasing perception noise and the thresholds $f$. It can be understood that the ability of the shepherd to guide the sheep deteriorates under larger noises, and instability is nearly reached. 

From this evaluation, it is interesting to demonstrate the robustness of the shepherding model against both sources of noise when the testing scenarios at $\alpha_2$, $\alpha_3$, and $\lambda_4$ could still have a 20\% minimum success rate. Regarding the changes of the threshold ($f$), when the noises increase, it is logical that this threshold should be increased to prefer the driving behaviour and reduce the collecting behaviour. Under large noises, there appears more sheep going out of the mass, and the shepherd might perform the collecting behaviour continuously, and then there is no chance to drive the sheep towards the target. This is a possible reason to see the task collapsing at the larger noise and the smaller threshold radius.

\section{Conclusion and Future Work}\label{sect:conclusions}
In this paper, we evaluate the performance of a shepherd guiding a swarm under actuation and perception noises. With 300 random episodes for 343 setups, the obtained results show that stability in performance is reached and maintained after the first 150  episodes at noise levels not exceeding the high level identified for both actuation and perception. When the noises are at high levels, the stability breaks down, and then the reliability of our ability to estimate the performance is very likely to become imprecise. After validating stability, a valuable point is drawn that the actuation noise has a higher impact on the performance of the shepherd than perception noise. The performance of the shepherd deteriorates earlier at a higher level of actuation noise, though this noise\textquoteright s value is less than one tens of the perception noise causing the same level of degradation in performance. 

Additionally, when the perception noise increases and the actuation noise is low, the lower collecting frequency leads to higher success rate. In contrast, when the actuation noise is higher and the perception noise is low, the higher collecting frequency contributes to higher success rate. These interesting results show promising evidences in order to design an adaptive behaviour controller, which allows the threshold $f$ to switch between the two collecting and driving behaviours to be adjusted, improving the performance of the shepherd under these noises. Our future work attempts to design this controller.
\bibliographystyle{IEEEtran}
\bibliography{references}

\begin{thebibliography}{10}
\providecommand{\url}[1]{#1}
\csname url@samestyle\endcsname
\providecommand{\newblock}{\relax}
\providecommand{\bibinfo}[2]{#2}
\providecommand{\BIBentrySTDinterwordspacing}{\spaceskip=0pt\relax}
\providecommand{\BIBentryALTinterwordstretchfactor}{4}
\providecommand{\BIBentryALTinterwordspacing}{\spaceskip=\fontdimen2\font plus
\BIBentryALTinterwordstretchfactor\fontdimen3\font minus
  \fontdimen4\font\relax}
\providecommand{\BIBforeignlanguage}[2]{{%
\expandafter\ifx\csname l@#1\endcsname\relax
\typeout{** WARNING: IEEEtran.bst: No hyphenation pattern has been}%
\typeout{** loaded for the language `#1'. Using the pattern for}%
\typeout{** the default language instead.}%
\else
\language=\csname l@#1\endcsname
\fi
#2}}
\providecommand{\BIBdecl}{\relax}
\BIBdecl

\bibitem{martinez2007motion}
S.~Martinez, J.~Cortes, and F.~Bullo, ``Motion coordination with distributed
  information,'' \emph{IEEE Control Systems Magazine}, vol.~27, no.~4, pp.
  75--88, 2007.

\bibitem{carelli2006centralized}
R.~Carelli, C.~De~la Cruz, and F.~Roberti, ``Centralized formation control of
  non-holonomic mobile robots,'' \emph{Latin American applied research},
  vol.~36, no.~2, pp. 63--69, 2006.

\bibitem{oh2017bio}
H.~Oh, A.~R. Shirazi, C.~Sun, and Y.~Jin, ``Bio-inspired self-organising
  multi-robot pattern formation: A review,'' \emph{Robotics and Autonomous
  Systems}, vol.~91, pp. 83--100, 2017.

\bibitem{strombom2018robot}
D.~Str{\"o}mbom and A.~J. King, ``Robot collection and transport of objects: A
  biomimetic process,'' \emph{Frontiers in Robotics and AI}, vol.~5, p.~48,
  2018.

\bibitem{nalepka2019practical}
P.~Nalepka, R.~W. Kallen, A.~Chemero, E.~Saltzman, and M.~J. Richardson,
  ``Practical applications of multiagent shepherding for human-machine
  interaction,'' in \emph{International Conference on Practical Applications of
  Agents and Multi-Agent Systems}.\hskip 1em plus 0.5em minus 0.4em\relax
  Springer, 2019, pp. 168--179.

\bibitem{cohen2014galvanotactic}
D.~J. Cohen, W.~J. Nelson, and M.~M. Maharbiz, ``Galvanotactic control of
  collective cell migration in epithelial monolayers,'' \emph{Nature
  materials}, vol.~13, no.~4, p. 409, 2014.

\bibitem{strombom2014solving}
D.~Str{\"o}mbom, R.~P. Mann, A.~M. Wilson, S.~Hailes, A.~J. Morton, D.~J.
  Sumpter, and A.~J. King, ``Solving the shepherding problem: heuristics for
  herding autonomous, interacting agents,'' \emph{Journal of the royal society
  interface}, vol.~11, no. 100, p. 20140719, 2014.

\bibitem{reynolds1987flocks}
C.~W. Reynolds, \emph{Flocks, herds and schools: A distributed behavioral
  model}.\hskip 1em plus 0.5em minus 0.4em\relax ACM, 1987, vol.~21, no.~4.

\bibitem{kennedy1995particle}
J.~Kennedy and R.~Eberhart, ``Particle swarm optimization,'' in \emph{Neural
  Networks, 1995. Proceedings., IEEE International Conference on},
  vol.~4.\hskip 1em plus 0.5em minus 0.4em\relax IEEE, 1995, pp. 1942--1948.

\bibitem{eberhart1995new}
R.~Eberhart and J.~Kennedy, ``A new optimizer using particle swarm theory,'' in
  \emph{Micro Machine and Human Science, 1995. MHS'95., Proceedings of the
  Sixth International Symposium on}.\hskip 1em plus 0.5em minus 0.4em\relax
  IEEE, 1995, pp. 39--43.

\bibitem{schultz1996roboshepherd}
A.~Schultz, J.~J. Grefenstette, and W.~Adams, ``Roboshepherd: Learning a
  complex behavior,'' \emph{Robotics and Manufacturing: Recent Trends in
  Research and Applications}, vol.~6, pp. 763--768, 1996.

\bibitem{lien2004shepherding}
J.-M. Lien, O.~B. Bayazit, R.~T. Sowell, S.~Rodriguez, and N.~M. Amato,
  ``Shepherding behaviors,'' in \emph{IEEE International Conference on Robotics
  and Automation, 2004. Proceedings. ICRA'04. 2004}, vol.~4.\hskip 1em plus
  0.5em minus 0.4em\relax IEEE, 2004, pp. 4159--4164.

\bibitem{bennett2012comparative}
B.~Bennett and M.~Trafankowski, ``A comparative investigation of herding
  algorithms,'' in \emph{Proc. Symp. on Understanding and Modelling Collective
  Phenomena (UMoCoP)}, 2012, pp. 33--38.

\bibitem{nguyen2019deep}
H.~T. Nguyen, T.~D. Nguyen, M.~Garratt, K.~Kasmarik, S.~Anavatti, M.~Barlow,
  and H.~A. Abbass, ``A deep hierarchical reinforcement learner for aerial
  shepherding of ground swarms,'' in \emph{International Conference on Neural
  Information Processing}.\hskip 1em plus 0.5em minus 0.4em\relax Springer,
  2019, pp. 658--669.

\bibitem{nguyen2019apprenticeship}
H.~T. Nguyen, M.~Garratt, L.~T. Bui, and H.~Abbass, ``Apprenticeship learning
  for continuous state spaces and actions in a swarm-guidance shepherding
  task,'' in \emph{2019 IEEE Symposium Series on Computational Intelligence
  (SSCI)}.\hskip 1em plus 0.5em minus 0.4em\relax IEEE, 2019, pp. 102--109.

\bibitem{gee2019transparent}
A.~Gee and H.~Abbass, ``Transparent machine education of neural networks for
  swarm shepherding using curriculum design,'' \emph{arXiv preprint
  arXiv:1903.09297}, 2019.

\bibitem{clayton2019machine}
N.~R. Clayton and H.~Abbass, ``Machine teaching in hierarchical genetic
  reinforcement learning: Curriculum design of reward functions for swarm
  shepherding,'' \emph{arXiv preprint arXiv:1901.00949}, 2019.

\end{thebibliography}
\end{document}